\documentstyle[12pt,psfig,worldsci]{article}
\begin{document}

\def\etal{{\it et al.}}
\def\ie{{\it i.e.}}
\def\eg{{\it e.g.}}
\def\wt{\widetilde}
\def\mpl{M_P}
\def\g{\ifmath{\gamma}}
\def\gev{{\rm GeV}}
\def\rta{\rightarrow}
\def\bp{\overline p}
\def\etmiss{E_T^{miss}}
\def\delr{\Delta R_{l,j}^{min}}
\def\china{\widetilde\chi^0_1}
\def\chinb{\widetilde\chi^0_2}
\def\chinc{\widetilde\chi^0_3}
\def\chind{\widetilde\chi^0_4}
\def\chini{\widetilde\chi^0_i}
\def\chinj{\widetilde\chi^0_j}
\def\chipa{\widetilde\chi^+_1}
\def\chipb{\widetilde\chi^+_2}
\def\chima{\widetilde\chi^-_1}
\def\chimb{\widetilde\chi^-_2}
\def\chipi{\widetilde\chi^+_i}
\def\chipmi{\widetilde\chi^\pm_i}
\def\chipj{\widetilde\chi^+_j}
\def\chimj{\widetilde\chi^-_j}
\def\chitil{\widetilde\chi}
\def\chitili{\widetilde\chi_i}
\def\chitilj{\widetilde\chi_j}
\def\mchitil{M_{\widetilde\chi}}
\def\mchipa{M_{\widetilde\chi^+_1}}
\def\mchipb{M_{\widetilde\chi^+_2}}
\def\mchina{M_{\widetilde\chi^0_1}}
\def\mchinb{M_{\widetilde\chi^0_2}}
\def\mchinc{M_{\widetilde\chi^0_3}}
\def\mchind{M_{\widetilde\chi^0_4}}
\def\mchini{M_{\widetilde\chi^0_i}}
\def\hpm{H^\pm}
\def\hone{H^0}
\def\htwo{h^0}
\def\hthree{A^0}
\def\mhpm{m_{H^\pm}}
\def\mhp{m_{H^+}}
\def\mhone{m_{H^0}}
\def\mhtwo{m_{h^0}}
\def\mhthree{m_{A^0}}
\def\mhl{m_{h^0}}
\def\mhh{m_{H^0}}
\def\mha{m_{A^0}}
\def\hl{h^0}
\def\hh{H^0}
\def\mz{m_Z}
\def\mw{m_W}
\def\ifmath#1{\relax\ifmmode #1\else $#1$\fi}
\def\half{\ifmath{{\textstyle{1 \over 2}}}}
\def\quarter{\ifmath{{\textstyle{1 \over 4}}}}
\def\sixth{\ifmath{{\textstyle{1 \over 6}}}}
\def\third{\ifmath{{\textstyle{1 \over 3}}}}
\def\twothirds{{\textstyle{2 \over 3}}}
\def\fourth{\ifmath{{\textstyle{1\over 4}}}}
\def\msq{M_{\widetilde q}}
\def\sq{\widetilde q}
\def\sql{\widetilde q_L}
\def\sqr{\widetilde q_R}
\def\mgl{M_{\widetilde g}}
\def\gl{\widetilde g}
\def\gp{g^{\prime}}
\def\lsim{\mathrel{\raise.3ex\hbox{$<$\kern-.75em\lower1ex\hbox{$\sim$}}}}
\def\gsim{\mathrel{\raise.3ex\hbox{$>$\kern-.75em\lower1ex\hbox{$\sim$}}}}

\pagestyle{empty}

%\begin{titlepage}
\begin{flushright}
CERN-TH/95-249\\
SCIPP 95/43 \\
hep--ph/9510412 \\
\end{flushright}
\vskip 0.3in
\begin{center}
{\Large\bf Low-Energy Supersymmetry: \\ Prospects and
Challenges}\\
\vspace{0.5cm}
Howard E. Haber\footnote{Permanent address:
Santa Cruz Institute for Particle Physics, University of California,
Santa Cruz, CA 94064  USA.}\\
\vspace{0.5cm}
CERN, TH Division, CH--1211 Geneva 23, Switzerland\\
\vskip1.3cm
{\bf Abstract}

\begin{quote}
An introduction to the
minimal supersymmetric Standard Model (MSSM) is given.
The motivation for ``low-energy'' supersymmetry is reviewed, and the
structure of the MSSM is outlined.
In its most general form, the MSSM can be viewed as a low-energy
effective theory parametrized by a set of arbitrary
soft-supersymmetry-breaking parameters.  A variety of techniques for
reducing the parameter freedom of the MSSM are surveyed.
The search for supersymmetry
below and above the threshold for supersymmetric particle production
presents a challenging task for experimentalists
at present and future colliders.
\end{quote}
\vskip1.cm
Invited Talk at the\\
International Workshop on Elementary Particle Physics,
Present and Future \\
Valencia, Spain, 5--9 June 1995\\
\end{center}
\vfill
\begin{flushleft}
CERN-TH/95-249\\
October 1995 \\
\end{flushleft}

%\end{titlepage}
%
\setcounter{footnote}{1}
\pagestyle{plain}
\newpage
%
% BODY
\section{Motivation for Low-Energy Supersymmetry}

The Standard Model of particle physics provides an extremely successful
description of all particle physics phenomena accessible to present
day accelerators.  No unambiguous experimental
deviation from the Standard Model have yet been confirmed.
However, theorists strongly believe that the success of the Standard
Model will not persist to higher energy scales.  This belief arises
{}from attempts to embed the Standard Model in a more fundamental
theory.  We know that the Standard Model cannot be the ultimate
theory, valid to arbitrarily high energy scales.  Even in the absence
of grand unification of strong and electroweak forces at a very high
energy scale, it is clear that the Standard Model must be
modified  to incorporate the effects of gravity at the Planck
scale ($\mpl\simeq 10^{19}$~GeV).  In this context, it is a mystery
why the ratio $\mw/\mpl\simeq 10^{-17}$ is so small.  This is called
the hierarchy problem \cite{4a,4b,Susskind84}.  Moreover,
in the Standard Model, the scale of
the electroweak interactions derives from an elementary scalar field
which acquires a vacuum expectation value of $v=2\mw/g=246$~GeV.
However, if one couples a theory of scalar particles to new physics at
some arbitrarily high scale $\Lambda$, radiative corrections to the
scalar squared-mass are of ${\cal O}(\Lambda^2)$, due to the quadratic
divergence in the scalar self-energy (which indicates quadratic
sensitivity to the largest energy scale, $\Lambda$,  in the theory). Thus, the
``natural'' mass for any scalar particle is $\Lambda$ (which is
presumably equal to $\mpl$).  Of course,
in order to have a successful electroweak theory, the Higgs mass must
be of order the electroweak scale.  The fact that the Higgs mass
must not be equal to its natural value of $\mpl$ is called the
``naturalness'' problem \cite{thooft}.  That is,
the bare Higgs squared-mass
and the squared-mass shift arising from the radiative corrections
are both expected to take on their natural values
of ${\cal O}(\mpl^2)$.  To end up with the physical Higgs
squared-mass of  ${\cal O}(\mw^2)$, which is 34 orders of magnitude
smaller than $\mpl^2$, requires a miraculous cancelation
(or ``fine-tuning'') among the parameters of the fundamental theory.

It is instructive to consider the following historical
precedent.
%\footnote{I thank H. Murayama for bringing this argument to
%to my attention.}
In the 1920's, quantum mechanics became the successful standard
model of fundamental physics.  But, this theory also possessed
a disturbing hierarchy problem: why is $m_e/\mpl\simeq{\cal O}
(10^{-22})$?  A calculation of the electron self-energy using
non-relativistic perturbation theory [see Fig.~1(a)]
yields a mass shift that
is linearly divergent.  That is, the natural value for $m_e$ is
the high energy scale $\Lambda$.  This behavior is not surprising.
After all, classically, the self-energy of an electron of radius
$r$ is $e^2/r$ which diverges linearly as $1/r\to\infty$.
The linear divergence persists in the relativistic single-electron
quantum theory.  How is this naturalness problem solved in quantum
electrodynamics?  The solution is remarkable.  Invent a new
symmetry called charge conjugation invariance (C).  Now, double
the known particle spectrum: for every particle, introduce a partner
called an ``antiparticle''.  The C symmetry guarantees that the
antiparticle has the same mass and interaction strength as its
partner.  Now, let us reconsider the perturbation theory computation
of the electron self-energy.  Now, there is a second diagram to
consider [see Fig.~1(b)], in which $e^+e^-\gamma$ is created from
the vacuum, the $e^+\gamma$ annihilates the incoming electron, while
the $e^-$ just created continues to propagate.  In old-fashioned
time ordered perturbation theory, both time orderings must be included
as shown in Fig.~1.  Due to the C symmetry, the leading linear
divergence cancels between the two graphs, leaving a logarithmic
divergence.  The mass shift of the electron
is thus proportional to $e^2m_e\ln\Lambda$.
The naturalness problem is solved, since even for $\Lambda=\mpl$, the
radiative correction to the electron mass is of the same order as $m_e$.
Of course, antiparticles were not invented to solve the naturalness
problem of the single-electron quantum
theory.  Nevertheless, the cancelation
of the linear divergence in Dirac's theory of electrons and positrons,
which was discovered by Weisskopf in 1934 \cite{Weisskopf}, was regarded
as an important advance in the development
of quantum electrodynamics \cite{earlyhistory}.

\begin{figure}[ht]
\vspace*{1pc}
\centering
\centerline{\psfig{file=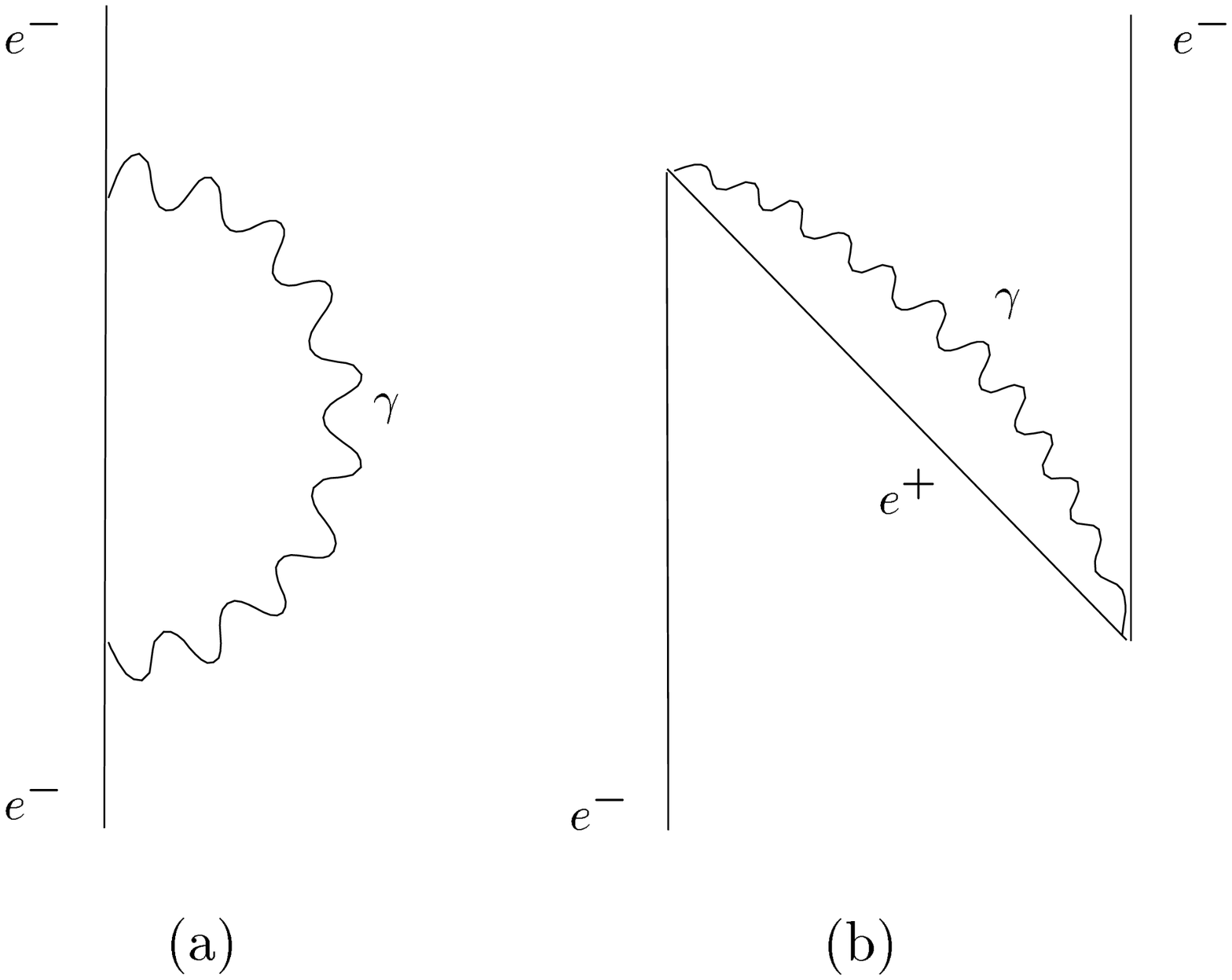,height=6cm}}
\centerline{Figure 1}
\end{figure}

To solve the naturalness problem of electroweak theory, we mimic
the steps just outlined.  In this case, we ``invent''
a new symmetry called supersymmetry, which transforms fermions into
bosons and vice versa.  Next, we
double the particle spectrum: for each particle we introduce a
superpartner which differs in spin by half a unit.
As in Dirac's theory of electrons and positrons,
the quadratic divergence of the scalar squared-mass is exactly
cancelled when the virtual exchange of superpartners is added to
the contributions of the Standard Model.  Thus in a supersymmetric
theory, the radiative corrections to the masses of both fermions and
bosons are at most logarithmically sensitive to the high energy scale
$\Lambda$.  Of course, the historical precedent does not provide an
exact analogy.  Because CPT symmetry in quantum field
theory must be exactly conserved, antiparticles
must be mass-degenerate with their particle partners.  In contrast,
supersymmetry cannot be an exact symmetry of nature, since
experimental data imply that supersymmetric particles are not mass degenerate
with their partners.  Nevertheless, if the scale of supersymmetry
breaking is of order 1 TeV or below, then the naturalness problem of
the Standard Model would be resolved.  In such
theories of ``low-energy'' supersymmetry, the
supersymmetry breaking scale is tied to the scale of
electroweak symmetry breaking \cite{4a,4b,Susskind84}.

In addition to providing a potential solution of the naturalness
problem of the Standard Model, supersymmetry provides an
attractive theoretical framework that may permit
the consistent unification of particle physics and gravity
\cite{Nilles84,Nath84,Green87}.  It therefore deserves serious
consideration as a theory of fundamental particle interactions.

\section{The Minimal Supersymmetric Standard Model (MSSM)}

The minimal supersymmetric extension of the Standard Model (MSSM)
consists of taking the Standard Model
and adding the corresponding supersymmetric partners \cite{Haber85}.
In addition, the MSSM contains two hypercharge $Y=\pm 1$ Higgs
doublets, which is the minimal structure for the Higgs sector of an
anomaly-free supersymmetric extension of the Standard Model.
The supersymmetric structure of the theory also requires (at least) two
Higgs doublets to generate mass for both ``up''-type and ``down''-type
quarks (and charged leptons) \cite{Inoue82,Gunion86}.
All renormalizable supersymmetric interactions
consistent with (global) $B-L$
 conservation ($B=$baryon number and
$L=$lepton number) are included.  Finally, the most general
soft-supersymmetry-breaking terms are added \cite{Girardello82}.
If supersymmetry is relevant
for explaining the scale of electroweak interactions, then the
mass parameters associated with the soft-supersymmetry-breaking
terms must be of order 1~TeV or below \cite{Barbieri88}.
%Some bounds on these parameters exist due to the absence
%of supersymmetry particle production at current accelerators, as well
%as the absence of any evidence for virtual supersymmetric particle
%exchange in a variety of Standard Model processes \cite{Bertolini91}.
Some bounds on these parameters exist due to the absence of
supersymmetric particle production at current accelerators; see
\cite{pdg} for a complete listing of supersymmetric particle
mass limits.  Additional constraints
arise from limits on the contributions of virtual supersymmetric
particle exchange to a variety of Standard Model processes \cite{Bertolini91}.
The impact of precision electroweak measurements at LEP and SLC
on the MSSM parameter space is discussed briefly in section~4.

As a consequence of $B-L$ invariance, the MSSM possesses a discrete
$R$-parity invariance, where
${R=(-1)^{3(B\hbox{--}L)+2S}}$
for a particle of spin $S$ \cite{Fayet77}.
Note that this formula implies that all the ordinary Standard Model
particles have even $R$-parity, whereas the corresponding
supersymmetric partners have odd $R$-parity.
The conservation of $R$-parity in scattering
and decay processes has a crucial impact on supersymmetric
phenomenology.  For example, starting from an initial state involving
ordinary ($R$-even) particles, it follows that
supersymmetric particles must be
produced in pairs.  In general, these particles are highly unstable
and decay quickly into lighter states.  However, $R$-parity invariance
also implies that
the lightest supersymmetric particle (LSP) is absolutely
stable, and must eventually be produced at the end of a decay chain
initiated by the decay of
a heavy unstable supersymmetric particle.
In order to be
consistent with cosmological constraints, the LSP is almost certainly
electrically and color neutral \cite{Ellis84}.
Consequently, the LSP is weakly-interacting in ordinary matter,
\ie\ it behaves like a heavy stable neutrino
and will escape detectors without being directly observed.
Thus, the canonical signature for $R$-parity conserving
supersymmetric theories
is missing (transverse)
energy, due to the escape of the LSP.
Some model builders attempt to relax the assumption of
$R$-parity conservation.  Models of the type must break $B-L$
and are therefore strongly constrained by experiment \cite{Dimopoulos90}.
In such models, the LSP is unstable and
supersymmetric particles can be singly produced and destroyed in
association with $B$ or $L$ violation.  These features
lead to a phenomenology of broken-$R$-parity
models that is very different from that of the MSSM.

%In the MSSM, supersymmetry breaking is accomplished by including the
%soft-supersymmetry breaking terms mentioned above.  These terms
%parametrize our ignorance of the fundamental mechanism of
%supersymmetry breaking.
%If this breaking occurs spontaneously,
%then (in the absence of supergravity) a massless Goldstone fermion
%called the {\it goldstino} ($\widetilde G$) must exist.  The
%goldstino would then be the LSP and could play an important role
%in supersymmetric phenomenology \cite{Fayet79}.
%\refmark\fayet\
%In models that
%incorporate supergravity, this picture changes.  If supergravity
%is spontaneously broken, the goldstino is absorbed (``eaten'')
%by the {\it gravitino} ($g_{3/2}$),
%the spin-3/2 partner of the graviton \cite{Deser77}.
%\refmark\superhiggs\
%By this super-Higgs mechanism, the gravitino acquires a mass
%($m_{3/2}$). In many models, the gravitino mass is of order the electroweak
%symmetry breaking scale, while its couplings are
%gravitational in strength \cite{Nilles84,Lahanas87}.
%\refmark\nilles\
%Such a gravitino would play no role in supersymmetric
%phenomenology at colliders.

The parameters of the MSSM are conveniently described by considering
separately the
supersymmetry-conserving sector and
the supersymmetry-breaking sector.
Supersymmetry breaking is accomplished by including the
most general set of
soft-supersymmetry breaking terms;  these terms
parametrize our ignorance of the fundamental mechanism of
supersymmetry breaking.
%INSERT 3
A careful discussion of the conventions used in defining the MSSM
parameters can be found in  \cite{Haber92}.
Among the parameters of the supersymmetry conserving
sector are: (i)~gauge couplings: $g_s$, $g$, and $\gp$, corresponding
to the Standard Model gauge group SU(3)$\times$SU(2)$\times$U(1)
respectively; (ii)~Higgs Yukawa couplings: $\lambda_e$, $\lambda_u$, and
$\lambda_d$ (which are $3\times 3$ matrices in flavor space); and
(iii)~a supersymmetry-conserving Higgs mass parameter $\mu$.  The
supersymmetry-breaking sector contains the following set of parameters:
(i)~gaugino Majorana masses $M_3$, $M_2$ and $M_1$ associated with
the SU(3), SU(2), and U(1) subgroups of the Standard Model;
(ii)~scalar mass matrices for the squarks and sleptons;
(iii)~Higgs-squark-squark trilinear interaction terms (the so-called
``$A$-parameters'') and corresponding terms involving the sleptons; and
(iv)~three scalar Higgs mass parameters---two
diagonal and one off-diagonal mass terms for the two Higgs doublets.
These three mass parameters can be re-expressed in terms of the two
Higgs vacuum expectation values, $v_1$ and $v_2$, and one physical Higgs
mass.  Here, $v_1$ ($v_2$) is the vacuum
expectation value of the Higgs field which couples exclusively
to down-type (up-type) quarks and leptons.  Note that $v_1^2+v_2^2=
(246~{\rm GeV})^2$ is fixed by the $W$ mass,
while the ratio
\begin{equation} \label{eqtanbeta}
\tan \beta = v_2/v_1
\end{equation}
is a free parameter of the model.

The supersymmetric constraints imply that the MSSM Higgs sector is
automatically CP-conserving (at tree-level).  Thus,
$\tan\beta$ is a real parameter (conventionally chosen to be positive),
and the physical neutral Higgs scalars are CP-eigenstates.
Nevertheless, the
MSSM does contain a number of possible new sources of CP violation.
For example, gaugino mass parameters, the $A$-parameters,
and $\mu$ may be
complex.
Some combination of these complex phases
must be less than of order
$10^{-2}$--$10^{-3}$
(for a supersymmetry-breaking scale of 100 GeV) to avoid generating electric
dipole moments for the neutron, electron, and atoms
in conflict with observed data \cite{Fischer92}.
However, these complex phases
have little impact on the direct searches
for supersymmetric particles, and are usually
ignored in experimental analyses.

Before describing the supersymmetric particle sector, let us
consider the Higgs sector of the MSSM \cite{Gunion90}.
There are five physical Higgs particles in this model: a charged Higgs
pair ($\hpm$), two CP-even neutral Higgs bosons (denoted by $\hl$
and $\hh$ where $\mhl\leq\mhh$) and one CP-odd neutral
Higgs boson ($\hthree$).  The properties of the Higgs sector are
determined by the Higgs potential which is made up of quadratic terms
[whose squared-mass coefficients were mentioned above eq.~(\ref{eqtanbeta})]
and quartic interaction terms.  The strengths of the interaction terms
are directly related to the gauge couplings by supersymmetry.
As a result,
$\tan\beta$ [defined in eq.~(\ref{eqtanbeta})]
and one Higgs mass determine: the Higgs spectrum,
an angle
$\alpha$ [which indicates the amount of
mixing of the original $Y=\pm 1$ Higgs doublet
states in the physical CP-even
scalars], and
the Higgs boson couplings.  When one-loop radiative corrections
are incorporated, additional parameters of the supersymmetric model
enter via virtual loops.  The impact of these corrections
can be significant \cite{Haber91,Okada91}.
For example, at tree-level, the MSSM predicts
 $\mhl\leq m_Z$ \cite{Inoue82,Gunion86}.  If true,
this would imply that experiments to be performed at
LEP-2 operating at its
maximum energy and luminosity would
rule out the MSSM if $\hl$ were not found.
However, this Higgs mass bound can be violated when the
radiative corrections are incorporated.  For example, in  \cite{Haber91},
the following approximate upper bound was obtained for
$\mhl$ (assuming $\mha>\mz$)
in the limit of $\mz\ll m_t\ll M_{\widetilde t}$ [where
top-squark ($\widetilde t_L$--$\widetilde t_R$) mixing is neglected]
%\lefteqnside=0pt
\begin{equation} \label{eqii}
\mhl^2  \lsim   \mz^2+ {3g^2\mz^4\over 16
\pi^2\mw^2}
\Biggl\{\left[{2m_t^4 -m_t^2\mz^2\over\mz^4}\right]
\ln\left({M_{\widetilde t}^2\over m_t^2}\right)
+{m_t^2\over 3 m_Z^2}\Biggr\}\,.
\end{equation}
More refined computations \cite{r1}
(which include the effects of
top-squark mixing at one-loop,
renormalization group improvement, and the
leading two-loop contributions) yield $\mhl\lsim 125$~GeV
for $m_t=175$~GeV and a top-squark mass of $M_{\widetilde t}=1$~TeV.
Clearly, the radiative corrections to the Higgs masses
have a significant impact
on the search for the MSSM Higgs bosons at LEP \cite{altarelli95}.

Consider next the
supersymmetric particle sector of the MSSM.
The {\it gluino} is the
color octet Majorana fermion partner of the gluon
with mass $M_{\widetilde g}=|M_3|$.
The supersymmetric partners of the electroweak gauge
and Higgs bosons (the {\it gauginos} and {\it higgsinos})
can mix.  As a result,
the physical mass eigenstates are model-dependent linear combinations
of these states, called {\it charginos} and {\it neutralinos}, which
are obtained by diagonalizing the corresponding mass matrices.
The chargino mass matrix depends on $M_2$, $\mu$, $\tan\beta$ and
$m_W$ \cite{Explicitforms}.  The corresponding chargino mass eigenstates
are denoted by $\chipa$ and $\chipb$,
with masses
\begin{eqnarray} \label{chimasses}
M^2_{\chipa,\chipb} & = &
\half\Biggl\{|\mu|^2+|M_2|^2+2m_W^2\mp
\Biggl[
\left(|\mu|^2+|M_2|^2+2m_W^2\right)^2 \nonumber \\
&&
-4|\mu|^2|M_2|^2-4m_W^4\sin^2 2\beta
+8m_W^2\sin 2\beta\,{\rm Re}(\mu M_2)
\Biggr]^{1/2}\Biggr\}\,,
\end{eqnarray}
where the states are ordered such that $\mchipa \leq \mchipb$.
If CP-violating effects are ignored (in which case, $M_2$ and
$\mu$ are real parameters), then one can choose a convention where
$\tan\beta$ and $M_2$ are positive.
(Note that the relative sign of $M_2$ and $\mu$ is meaningful.
The sign of $\mu$ is convention-dependent; the
reader is warned that both sign conventions
appear in the literature.)
The sign convention for $\mu$ implicit in eq.~(\ref{chimasses})
is used by the LEP collaborations \cite{Decamp92}
in their plots of exclusion contours in the $M_2$ {\it vs.} $\mu$
plane derived from the non-observation of $Z\rightarrow\chipa\chima$.
The $4\times 4$ neutralino mass matrix depends on $M_1$, $M_2$, $\mu$,
$\tan\beta$, $m_Z$, and the weak mixing angle
$\theta_W$ \cite{Explicitforms}.
The corresponding neutralino eigenstates are usually denoted by
$\chini$ ($i=1,\ldots 4$), according to the convention that
$\mchina\leq\mchinb\leq\mchinc\leq\mchind$.
Typically, $\china$ is the LSP.
%If a chargino
%or neutralino eigenstate approximates a particular gaugino or
%higgsino state, it may be convenient to use the corresponding
%nomenclature.  For example, if $M_1$ and $M_2$ are small compared to
%$m_Z$ (and $\mu$), then the lightest neutralino $\widetilde\chi_1^0$
%will be nearly a pure photino,
%$\widetilde\gamma$ (the supersymmetric partner of the photon).

It is common practice in the literature to reduce the
supersymmetric parameter freedom
by requiring that all three gaugino mass parameters
are equal at some grand unification scale.  Then,
at the electroweak scale, the gaugino mass parameters can be expressed in
terms of one of them (say, $M_2$) and the gauge coupling constants:
%The other
%two gaugino mass parameters are given by
\begin{equation} \label{eqmass3}
M_3=(g_s^2/g^2)M_2\ ,\qquad
M_1=(5g^{\prime\,2}/3g^2)M_2~.
\end{equation}
Having made this assumption, the chargino and neutralino masses and
mixing angles depend only on three unknown parameters: the gluino mass,
$\mu$, and $\tan\beta$.  However, the assumption of gaugino mass
unification could prove false and must eventually be tested experimentally.

The supersymmetric partners of the quarks and leptons are spin-zero
bosons:  the {\it squarks}, charged {\it sleptons}, and {\it sneutrinos}.
For a given fermion
$f$, there are two supersymmetric partners $\widetilde
f_L$ and $\widetilde f_R$ which are scalar partners of the
corresponding left and
right-handed fermion.  (There is no $\widetilde\nu_R$.)  However, in
general,
$\widetilde f_L$ and $\widetilde f_R$ are not mass-eigenstates since there
is $\widetilde f_L$-$\widetilde f_R$ mixing which is proportional in
strength to the corresponding element of the scalar
squared-mass matrix \cite{Ellis83}:
\begin{equation} \label{eqmass2LR}
M_{LR}^2={\cases{m_d(A_d-\mu\tan\beta),&for ``down''-type $f$\cr
       m_u(A_u-\mu\cot\beta),&for ``up''-type $f$,\cr}}
\end{equation}
where $m_d$ ($m_u$) is the mass of the appropriate
``down'' (``up'') type quark or lepton.
Here, $A_d$ and $A_u$ are (unknown)
soft-supersymmetry-breaking $A$--parameters and $\mu$ and $\tan\beta$
have been defined earlier.
The signs of the $A$~parameters are also
convention-dependent; see  \cite{Haber92}.
Due to the appearance
of the {\it fermion} mass in eq.~(\ref{eqmass2LR}),
 one expects $M_{LR}$ to be small
compared to the diagonal squark and slepton masses, with the possible
exception of the top-squark, since $m_t$ is large,
and the bottom-squark
and tau-slepton if $\tan\beta\gg1$.
The (diagonal) $L$ and $R$-type
squark and slepton masses are given by \cite{Nath84}
\begin{eqnarray}
 M^2_{\widetilde u_L}
	&=& M^2_{\wt Q}+m_u^2+m_Z^2\cos 2\beta(\half-
\twothirds \sin^2\theta_W)\label{eqmass2mubarL} \\
  M^2_{\widetilde u_R}
	&=& M^2_{\wt U}+m_u^2+\twothirds m_Z^2\cos 2\beta
\sin^2\theta_W\label{eqmass2mubarR} \\
  M^2_{\widetilde d_L}
	&=& M^2_{\wt Q}+m_d^2-m_Z^2\cos 2\beta(\half-\third
\sin^2\theta_W) \label{eqmass2dbarL} \\
   M^2_{\widetilde d_R}
	&=& M^2_{\wt D}+m_d^2-\third m_Z^2\cos 2\beta
\sin^2\theta_W \label{eqmass2dbarR} \\
   M^2_{\widetilde \nu}\;
	&=& M^2_{\wt L}+\half m_Z^2\cos 2\beta \label{eqmass2nubar} \\
   M^2_{\widetilde e_L}
	&=& M^2_{\wt L}+m_e^2-m_Z^2\cos 2\beta(\half-
\sin^2\theta_W) \label{eqmass2ebarL} \\
   M^2_{\widetilde e_R}
	&=& M^2_{\wt E}+m_e^2-m_Z^2\cos 2\beta
	\sin^2\theta_W\,. \label{eqmass2ebarR}
\end{eqnarray}
The soft-supersymmetry-breaking parameters: $M_{\wt Q}$,
$M_{\wt U}$, $M_{\wt D}$, $M_{\wt L}$, and $M_{\wt E}$ are
unknown parameters.  In the equations above, the notation of first
generation fermions has been used and generational indices have been
suppressed.  Further
complications such as intergenerational mixing are possible, although
there are some constraints from the nonobservation of flavor-changing
neutral currents (FCNC) \cite{recentworks}.

\section{Reducing the Supersymmetric Parameter  Freedom}

One way to guarantee the absence of significant FCNC's mediated
by virtual supersymmetric particle exchange is to posit that the
diagonal soft-supersymmetry-breaking scalar squared-masses are
universal in flavor space at some energy
scale (normally taken to be at or near
the Planck scale) \cite{4b,Dine93,r2}.
Renormalization
group evolution is used to determine the
low-energy values for the scalar
mass parameters listed above.  This assumption substantially reduces
the MSSM parameter freedom.  For example, supersymmetric grand unified
models with universal scalar masses at the Planck scale typically
give \cite{Drees95}
${M_{\wt L} \approx M_{\wt E} < M_{\wt Q} \approx}$
${M_{\wt U}
\approx M_{\wt D}}$
with the squark masses somewhere between a factor of 1--3 larger
than the slepton masses (neglecting generational
distinctions).  More specifically,
the first two generations are thought to be nearly degenerate in
mass, while
$M_{\wt Q_3}$ and $M_{\wt U_3}$ are typically reduced by a factor of 1--3
{}from the other soft-supersymmetry-breaking masses because
of renormalization effects due to the heavy top quark mass.
As a result, four flavors of squarks
(with two squark eigenstates per flavor)
 and $\wt b_R$ will be nearly mass-degenerate and somewhat heavier than six
flavors of
nearly mass-degenerate sleptons (with two per flavor for the charged
sleptons and one per flavor for the sneutrinos).
On the other hand, the
$\wt b_L$ mass and the diagonal
$\widetilde t_L$ and
$\widetilde t_R$ masses are reduced compared to the common squark mass of the
 first
two generations.  In addition, third generation squark masses and
tau-slepton masses are
sensitive to the strength of the respective
$\wt f_L$--$\wt f_R$ mixing as discussed below eq.~(\ref{eqmass2LR}).

Two additional theoretical frameworks are often introduced to reduce
further the MSSM parameter
freedom \cite{Nilles84,Nath84,Drees95,Arnowitt93}.
The first involves grand unified
theories (GUTs) and the desert hypothesis ({\it i.e.} no new physics
between the TeV-scale and the GUT-scale).
Perhaps one of the most compelling hints for low-energy supersymmetry
is the unification of SU(3)$\times$SU(2)$\times$U(1) gauge couplings
predicted by supersymmetric GUT models \cite{4b,rg}\
(with the supersymmetry breaking
scale of order 1 TeV or below).  The unification, which takes place at
an energy scale of order $10^{16}$~GeV, is quite robust (and depends weakly
on the details of the GUT-scale theory).  For example,
a recent analysis \cite{r7}
finds that supersymmetric GUT unification implies that
$\alpha_s(m_Z)=0.129\pm 0.010$, not including threshold corrections
due to GUT-scale particles (which could diminish the value
of $\alpha_s(m_Z)$).  This result is compatible with the world average of
$\alpha_s(m_Z)=0.117\pm 0.005$ \cite{pdg}.
In contrast, gauge coupling
unification in the simplest nonsupersymmetric GUT models fails by many
standard deviations \cite{r8}.  Grand unification can impose additional
constraints through the unification of Higgs-fermion Yukawa couplings
($\lambda_f$).  There is some evidence that $\lambda_b=\lambda_\tau$
leads to good low-energy phenomenology \cite{r9}, and an intriguing possibility
that in the MSSM (in the parameter regime where $\tan\beta\simeq m_t/m_b$)
$\lambda_b=\lambda_\tau=\lambda_t$ may be phenomenologically viable \cite{r10}.
However, such unification constraints are GUT-model dependent, and do not
address the origin of the first and second generation fermion masses and
the CKM mixing matrix.  Finally, grand unification imposes constraints on the
soft-supersymmetry-breaking parameters.  For example, gaugino mass
unification leads to the relations given in eq.~(\ref{eqmass3}).
Diagonal squark and slepton soft-supersymmetry-breaking scalar
masses may also be unified at the GUT scale
(analogous to the unification of Higgs-fermion Yukawa couplings).

In order to further reduce the number of independent soft-supersymmetry
breaking parameters (with or without grand unification), an additional
simplifying assumption is required.
In the minimal supergravity theory, the soft supersymmetry-breaking
parameters are often taken to have the following simple form.
Referring to the parameter list given above eq.~(1),
the Planck-scale values of the soft-supersymmetry-breaking terms
depend on the following minimal set of parameters:
(i) a universal gaugino mass $m_{1/2}$; (ii) a universal
diagonal scalar mass parameter $m_0$
[whose consequences were described at the beginning of this section];
(iii) a universal $A$-parameter, $A_0$; and (iv) three scalar Higgs
mass parameters---two common diagonal squared-masses given by $|\mu_0|^2+m_0^2$
and an off-diagonal squared-mass given by  $B_0\mu_0$
 (which defines the Planck-scale supersymmetry-breaking
parameter $B_0$), where $\mu_0$ is
the Planck-scale value of the $\mu$-parameter.  As before,
renormalization group evolution is used to compute the low-energy values
of the supersymmetry-breaking parameters and determines the
supersymmetric particle spectrum.  Moreover, in this approach,
electroweak symmetry breaking is induced radiatively if one of the
Higgs diagonal squared-masses is forced negative by the evolution.
This occurs in models with a large Higgs-top quark Yukawa
coupling ({\it i.e.} large $m_t$).  As a result, the two Higgs
vacuum expectation values (or equivalently, $m_Z$ and $\tan\beta$)
can be expressed as a function of the Planck-scale supergravity
parameters.  The simplest procedure \cite{Drees95,Arnowitt93}
is to remove $\mu_0$ and $B_0$
in favor of $m_Z$ and $\tan\beta$ (the
sign of $\mu_0$ is not fixed in this
process). In this case, the MSSM spectrum
and its interactions are determined by $m_0$, $A_0$, $m_{1/2}$,
$\tan\beta$, and the
sign of $\mu_0$  (in addition to the parameters of the Standard Model).
%Combining both grand unification and the minimal supergravity
%approach yield the most constrained version of the MSSM.
However, the minimal approach above is probably too restrictive.
Theoretical considerations suggest that the universality of Planck-scale
soft-supersymmetry breaking parameters is not generic \cite{r11}.
In the absence
of a fundamental theory of supersymmetry breaking, further progress will
require a detailed knowledge of the supersymmetric particle spectrum
in order to determine the nature of the Planck-scale parameters.

\section{Challenges for Supersymmetry Searches}

The verification of low-energy supersymmetry requires the discovery of the
supersymmetric particles.  Once superpartners are discovered, it is
necessary to test their detailed properties to verify the supersymmetric
nature of their interactions.  Furthermore, one can explicitly test
many of the additional theoretical assumptions of section~3
that were introduced to reduce the supersymmetric parameter freedom.

The search for supersymmetry at present and future colliders
falls into two distinct classes.  At colliders
whose energies lie below supersymmetric particle production threshold,
indirect effects of supersymmetry may be observable.  For example,
in the Higgs sector, if $\mha\gg\mhl$, then the properties of $\hl$
will be nearly indistinguishable from the Higgs boson of the minimal
Standard Model \cite{haberthomas}.  Small deviations from the Standard
Model Higgs sector could signal the existence of additional Higgs
states, as expected in the MSSM.
One can also search for deviations from Standard Model predictions due
to the effects of virtual supersymmetric particle exchange.
Such effects could be revealed in the measurement of
precision electroweak observables.  In both cases, one is fighting
the decoupling limit.  That is, in the limit that
soft-supersymmetry-breaking masses (collectively denoted by
$M_{\rm SUSY}$) become
large, the MSSM below supersymmetric threshold precisely reproduces the
predictions of the Standard Model.  At colliders whose energies lie
above supersymmetric particle production threshold, the direct effects
of supersymmetric production and decay are detectable.  In this case,
once superpartners are discovered, one must elucidate the details of
the low-energy supersymmetric theory.

The MSSM (with or without constraints imposed from the theory near the
Planck scale) provides a framework that can be tested by
precision electroweak data.  The level of accuracy of the measured
$Z$ decay observables at LEP and SLC is sufficient to test the
structure of the one-loop radiative corrections of the electroweak
model \cite{r3}, and is thus potentially sensitive to the virtual effects of
undiscovered particles.  Combining the most recent LEP and SLC
electroweak results \cite{r4} with the recent top-quark mass measurement
at the Tevatron \cite{r5}, a weak preference is
found \cite{r4,r4a} for a light
Higgs boson mass of order $m_Z$, which is consistent with the
MSSM Higgs mass upper bound noted in section~2.
Moreover, for $Z$ decay observables, the effects of virtual
supersymmetric particle exchange are suppressed by a factor
of $m_Z^2/M_{\rm SUSY}^2$, and therefore decouple
in the limit of large supersymmetric particle masses.
It follows that for $M_{\rm SUSY}^2\gg m_Z$ (in practice,
it is sufficient to have all supersymmetric particle masses above 200~GeV)
the MSSM yields an equally good fit to the precision electroweak data as
compared to the Standard Model fit.  On the other hand, there are a few
tantalizing hints in the data for deviations from Standard Model
predictions.  Indeed, if $R_b\equiv \Gamma(Z\rightarrow b\bar b)/
\Gamma(Z\rightarrow{\rm hadrons})$ is confirmed to lie above its
Standard Model prediction due to the presence of new physics, then
a plausible candidate for the new physics would be the MSSM with
some light supersymmetric particles ({\it e.g.} a light chargino and
top-squark and/or a light CP-odd scalar, $A^0$) close in mass to their
present LEP bounds \cite{r6,r6a}.  Such a scenario would be
tested by the search for supersymmetric particles at LEP-2 and the
Tevatron.

\def\eslt{E\llap/_T}
\def\etmiss{E\llap/_T}
\def\tg{\tilde g}
\def\tq{\tilde q}
\def\tl{\tilde \ell}
\def\tz{\widetilde\chi^0}
\def\tw{\widetilde\chi^\pm}
If low-energy supersymmetry exists, it should
be discovered at either upgrades of existing colliders or at the
LHC \cite{barklow,baer}.  Due to its mass reach,
the LHC is the definitive machine for discovering or
excluding low-energy supersymmetry.   Table 1 summarizes
the supersymmetry mass discovery potential for hadron colliders.
A variety of signatures are considered.
Many of the supersymmetry searches rely on the missing
energy signature as an indication of new physics beyond the
Standard Model.  Multi-leptonic signatures also play an
important role in supersymmetry searches at hadron colliders.
(Such signals can also be exploited in the search for
$R$-parity-violating low-energy supersymmetry.)
A comprehensive analysis can be found in \cite{baer}.

\begin{table*}[t]
\vspace*{-1.5pc}
\begin{center}
%\let\normalsize=\captsize
%\begin{minipage}{18.1cm}
\caption{Discovery reach of various options of future hadron
colliders \protect\cite{baertable}.
The numbers are subject to $\pm$15\% ambiguity. Also, the clean
trilepton signals are sensitive to other model parameters;
representative ranges from Ref.~\protect\cite{bcpt} are shown
where $|\mu |$ is typically much larger
than the soft-breaking electroweak gaugino masses.
For $\mu>0$, the leptonic decay of $\tz_2$ may be strongly suppressed
so that $3\ell$ signals may not be observable even if charginos
are just above the LEP bound.} %$M_{\tg}\sim 150$~GeV).}
\protect\label{discovery}
\vskip 0.5\baselineskip
%\footnotesize
\renewcommand\tabcolsep{3pt}
\renewcommand\arraystretch{1.3}
\small
\begin{tabular}{|l|cccccc|}
\hline
&Tevatron I&Tevatron II&Main Injector&Tevatron$^*$&DiTevatron&LHC\\[-2pt]
Signal&0.01~fb$^{-1}$&0.1~fb$^{-1}$&1~fb$^{-1}$&10~fb$^{-1}$&
1~fb$^{-1}$&10~fb$^{-1}$\\[-2pt]
&1.8~TeV&1.8~TeV&2~TeV&2~TeV&4~TeV&14~TeV\\
\hline
$\etmiss (\tq\!\gg\!\tg)$ & $\tg(150)$ & $\tg(210)$ & $\tg(270)$ &
$\tg(340)$ & $\tg(450)$ & $\tg(1300)$ \\
$\ell^\pm \ell^\pm (\tq\!\gg\!\tg)$ & --- & $\tg(160)$ & $\tg(210)$ &
$\tg(270)$ & $\tg(320)$ & $\tg(1000)$ \\
$all\to 3\ell (\tq\!\gg\!\tg)$ & --- & $\tg(150$-$180)$ &
$\tg(150$-$260)$ & $\tg(150$-$430)$ & $\tg(150$-$320)$ & \\
$\etmiss (\tq\!\sim\!\tg)$ & $\tg(220)$ & $\tg(300)$ & $\tg(350)$ &
$\tg(400)$ & $\tg(580)$ & $\tg(2000)$ \\
$\ell^\pm \ell^\pm (\tq \sim \tg)$ & --- & $\tg(180$-$230)$ & $\tg(325)$ &
$\tg(385$-$405)$ & $\tg(460)$ & $\tg(1000)$ \\
$all\to 3\ell (\tq\!\sim\!\tg)$ & --- & $\tg(240$-$290)$ &
$\tg(425$-$440)$ & $\tg(550)$ & $\tg(550)$ & $\gsim\!\tg(1000)$ \\
$\tilde{t}_1 \rightarrow c \tz_1$ & --- & $\tilde{t}_1(80$-$100)$ &
$\tilde{t}_1 (120)$ & & & \\
$\tilde{t}_1 \rightarrow b \tw_1$ & --- & $\tilde{t}_1(80$-$100)$ &
$\tilde{t}_1 (120)$ & & & \\
$\Theta(\tilde{t}_1 \tilde{t}_1^*)\rightarrow \gamma\gamma$ &
--- & --- & --- & --- & --- & $\tilde{t}_1 (250)$\\
$\tl \tl^*$ & --- & $\tl(50)$ & $\tl(50)$ & $\tl(100)$ & &
$\tl(250$-$300)$ \\
\hline
\end{tabular}
%\end{minipage}
\end{center}
\end{table*}

Suppose that a signal is observed in one of the expected channels.
This would not be
a confirmation of low-energy supersymmetry, unless
there is confirming evidence from other expected signatures.
This presents a formidable challenge to experimenters at the LHC.
Can they prove that a set of signatures of new physics is low-energy
supersymmetry?  Can they extract parameters of the supersymmetric
models with any precision and test the details of the theory?
These are questions that have only recently attracted seriously
study.  It is in this context that a future $e^+e^-$ collider (NLC)
can be invaluable.
If the lightest supersymmetric particles were
produced at LEP-2 or the NLC, precision measurements could begin
to map out in detail the parameter space of the supersymmetric
model.  In particular, beam polarization at the NLC provides an
critical tool for studying the relation between chirality and
the properties of supersymmetric particles \cite{japan}.
One can then begin to demonstrate that there is a correlation between the
left and right-handed electrons and their slepton partners as
expected in supersymmetry.  Moreover,
the determination of superparticle masses allows one to test theoretical
assumptions at various levels.
For example, the universality of slepton masses can
be tested at the 1\% level. More experimentally challenging is the test
of the GUT-relation among gaugino masses [eq.~(\ref{eqmass3})].
However, one can still test eq.~(\ref{eqmass3})
at the few percent level by combining slepton and chargino
signals, based on the measured masses and polarization dependence of the
cross sections.  See \cite{baer} for further details.
%The elucidation of the details of
%the light chargino and neutralino spectrum at the NLC could
%also play a pivotal role in untangling complex decay chains of
%heavier supersymmetric particles produced at the LHC.

Many theorists believe that the prospects for supersymmetry are
excellent.  Nevertheless,
the search for supersymmetry at future colliders may reveal
many surprises and raise new challenges for both theorists and
experimentalists.  If low-energy supersymmetry is discovered it will
have a profound effect on the development of 21st century theories of
fundamental particles and their interactions.

\begin{center}
\vskip 1pc
{\bf Acknowledgments}
\end{center}

I would like to thank Hitoshi Murayama for bringing to my attention the
naturalness problem of the single-electron quantum theory.
%argument that the antiparticles are responsible for eliminating the
%linear divergence in the fermion self-energy.
I am grateful to Jon Bagger, Vernon Barger,
Michael Barnett, Paul Langacker and Michael Peskin for
their comments on an earlier version of this manuscript.
Finally, I acknowledge the gracious hospitality and
support of Jos\'e Valle and his colleagues during my visit to Valencia.
This work was supported in part by the US Department of Energy.

%%%%%%%%%%%%%%%%%%%%%%%%%%%%%%%%%%%%%%%%%%%%%%%%%%%%%%%%%%%%%%%%%%%%%%%%


\begin{thebibliography}{99}
\bibitem{Explicitforms}%\REF\inos
Explicit forms for the chargino and neutralino mass matrices
can be found in Appendix~A of  \cite{Gunion86}; see also
 \cite{Haber92}.
\end{thebibliography}
\end{document}